\documentclass[pdflatex,sn-nature]{sn-jnl}

\usepackage{graphicx}
\usepackage[cmex10]{amsmath}
\usepackage{tikz}
\usetikzlibrary{arrows.meta, positioning, shapes.multipart, quotes}
\usepackage{array}
\usepackage{color}
\usepackage{amssymb}
\usepackage{balance}
\usepackage{booktabs}
\definecolor{mintbg}{rgb}{.63,.79,.95}
\colorlet{lightmintbg}{mintbg!40!white}
\usepackage{circuitikz}
\usepackage{float}
\usepackage{hyperref}
\usepackage{makecell}
\usepackage{lmodern}

\newcommand{\ii}{\text{i}} 

\begin{document}

\title[Solving Sudoku using oscillatory neural networks]{Solving Sudoku using oscillatory neural networks}

\author[1]{
    \fnm{Bram F.}
    \sur{Haverkort}
}
\equalcont{These authors contributed equally to this work.}
\author[1]{
    \fnm{Federico}
    \sur{Sbravati}
}
\equalcont{These authors contributed equally to this work.}
\author[1]{
    \fnm{Stefan}
    \sur{Porfir}
}
\author*[1]{
    \fnm{Aida}
    \sur{Todri-Sanial}
    \email{a.todri.sanial@tue.nl}
}

\affil[1]{
    \orgdiv{Integrated Circuits, Electrical Engineering},
    \orgname{Eindhoven University of Technology},
    \city{Eindhoven},
    \country{The Netherlands}%
}


\abstract{
    We explore the capabilities of physical computing with Oscillatory Neural Networks (ONN) to solve combinatorial optimization problems.
    To solve Sudokus with ONNs, we define a novel mapping strategy that utilizes the unique characteristics of the computation paradigm.
    The problem is encoded through a puzzle specific graph-embedding, which implements the constraints through different subgraphs.
    These subgraphs are then combined into a single \textit{adjacency matrix}, which allows the natural dynamics of the phases of coupled oscillators to find a solution to the puzzle.
    We model the phase dynamics of the ONN by means of the Kuramoto differential equation.
    This novel approach is then compared to the well-established iterative method to solve Sudoku already used in binary Hopfield networks (HNN).
    Solving optimization problems typically requires a large amount of energy to solve on conventional hardware.
    Therefore, we are motivated to explore the mapping of Sudoku from a theoretical point of view to establish the validity of this approach.
    The simulation results show that the novel ONN mapping outperforms the established HNN methodology.
}

\maketitle
\section{Introduction}\label{sec:introduction}
Since the rise of modern computing, the von Neumann architecture played an important role for shaping the way information is processed \cite{Backus1978}. 
However, as computing demands grow, the limitations of this architecture are an increasingly relevant topic \cite{Zou2021}. 
The main issue is due to the spatial separation between \textit{memory} and \textit{data processing units}, which imposes limitations on data throughput and energy efficiency \cite{ivanov2022neuromorphicartificialintelligencesystems}. 
The field of neuromorphic and physical computing aims to implement energy efficient hardware by taking inspiration from the parallel and event-driven nature of the brain \cite{todrisanial_lirmm-03022129}, \cite{markovic2020physics} and physical phenomena \cite{todri-sanial_computing_2024}.
From these fields, Oscillatory Neural Networks (ONNs) \cite{todri-sanial_computing_2024} have emerged as a biologically inspired computing paradigm that exploits the dynamics of coupled oscillators, while maintaining low-energy characteristics \cite{todrisanial_lirmm-03022129}.
\begin{figure}
    \centering
    \includegraphics[width=0.5\linewidth]{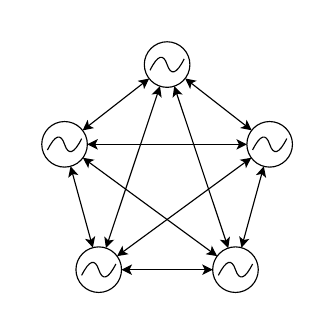}
    \caption{Example of an Oscillatory Neural Network, made up of five elements. 
    Each node represents the dynamical state of the phase of an oscillator.}
    \label{fig:example_onn}
\end{figure}
ONNs simulate the phase evolution of neural oscillations in the brain through a network of coupled oscillators.
The peculiarity of ONNs is that information is encoded in the phase relations between pairs of oscillators, which enables parallel computation of the states of the phases \cite{izhikevich2000computing}.
Contrary to binary approaches, such as the Hopfield Neural Network (HNN) \cite{Hopfield_1982}, this allows for the encoding of multiple states in one neuron \footnote{
    It is important to remark that depending on the application, the encoding to discrete steady-states is desirable in order to have unambiguous encoding within a certain tolerance.
}.
Fig. \ref{fig:example_onn} shows a five oscillator network.
%

Sudoku can be formulated as a constraint satisfaction problem \cite{simonis2005sudoku}, where each digit assignment must meet unique constraints across the grid. 
Traditional approaches to solving such problems include backtracking \cite{civicioglu2013backtracking} or more advanced techniques such as SAT solvers \cite{lynce2006sudoku}.
Oscillators have also been employed in convolutional neural network as attention layers to solve Sudoku \cite{miyato_artificial_2025}, but such approaches use training instead of directly exploiting the natural dynamics of the oscillators.
The existing literature has tackled solving Sudoku through Artificial Neural Networks, such as HNNs \cite{hopfield_searching_2008, mladenov2011solving}, allowing us to compare the performance of HNNs with our own approach using ONNs.

The goal of this paper is to define a mapping for ONNs to solve Sudoku by means of the natural dynamics of the coupled oscillator system.
Furthermore, our goal is to provide a comparison of the results with the classical binary HNN approach \cite{hopfield_searching_2008, mladenov2011solving}.
In section \ref{sec:background} we introduce the Kuramoto model, whose dynamics we exploit for the mapping and solution of Sudoku.
In section \ref{sec:methods} we discuss the way the Sudoku digits are mapped to phase values to make them compatible with the Kuramoto model, how the graph topology of the Sudoku constraints is related to the Weight Matrix used to couple the phases of the neurons in the Kuramoto model, how the synchronization order parameter\footnote{
    We do not quantify the energy or Hamiltonian of the system, because in the present mapping the connections are not Hermitian in general. 
    Therefore, the system cannot be trivially linked to the gradient descent of a potential energy.
    As a result the synchronization order parameter $\rho$ is a more appropriate quantity.
} is used in the context of this work, how the \textit{Time-to-Settle} metric used to assess the performance of the network to find a solution to the puzzle, and finally the benchmarking methodology for the dataset used in the paper.
We present the results in section \ref{sec:results} and discuss them in section \ref{sec:discussion}, after which we present the conclusions of the work in section \ref{sec:conclusions}.

\section{Background}\label{sec:background}

\subsection{Kuramoto model}
The model that describes the phase dynamics is the Kuramoto Model \cite{mori_dissipative_1998}: 
\begin{equation}
    \dot{\theta}_{i} = \omega^{(0)}_{i} - \sum_{j=1}^{N} J_{ij} \sin(\theta_{i} - \theta_{j}) \,\,\,,
\end{equation}\label{eq:kuramoto-model}
where $\theta_{i}$ is the phase of oscillator $i$ (in radians), $\dot{\theta}_{i}$ is the instantaneous rate of change of the phase (angular velocity) of the $i$-th oscillator, $\omega^{(0)}_{i}$ is the natural frequency of oscillator $i$, $N$ is the number of oscillators in the network, and $J_{ij}$ is the coupling strength between oscillators $i$ and $j$. 
The  Kuramoto model is all-to-all connected in its standard form, with a real and symmetric coupling matrix.
Since the puzzle possesses different topological requirements, we use a modified form with an ad-hoc topology, encoded by the connection matrix $J_{ij}$.
In general, the matrix $J_{ij}$ is complex and non-symmetric in the presented mapping.
We assume the natural frequency $\omega^{(0)}_{i}$  to be the same for each oscillator, and is thus neglected in the simulations\footnote{
    The reason why this is a physically reasonable assumption, is due to the fact that electronic oscillators can be calibrated to have the same natural frequency.
}.

\subsection{Sudoku puzzles}
Sudoku is a puzzle where the goal is to complete a partially filled square grid of digits according to three rules.
In its classical form it is a $9\times9$ grid, but it can be generalized to other sizes.
The linear size of the grid, in this instance $9$, is the same as the number of digits needed to solve the puzzle.
Each $3\times3$ sub cell must contain each digit from $1$ to $9$ exactly once, each row must contain each digit from $1$ to $9$ exactly once, and each column must contain each digit from $1$ to $9$ exactly once.
Typically, the pre-filled, or \textit{known}, cells are given in such a way that there is a unique solution to the puzzle.
Note that it is possible to have non-uniquely solvable Sudoku when the amount of known digits or their positions are such that multiple solutions are associated to the same puzzle. 
We try to avoid these kinds of puzzles in this paper.
We show an example of an unsolved Sudoku in Fig. \ref{fig:sudoku_9_9}.
\begin{figure}
    \centering
    \begin{tikzpicture}
  \def\cellsize{0.75cm}
  \def\tabledata{
    {1, ,7, , , , ,9, },
    { ,4,2, ,7,9,1, , },
    { , , , ,2, , , ,5},
    { ,3, ,2, ,5, , , },
    {5, , , , , , ,8, },
    { , ,4,6,9, ,3, , },
    { , ,1, , ,2,4, , },
    {4,2, , , , , , ,8},
    {3, , , ,6,7, , ,1},
  }

  \foreach \row [count = \i] in \tabledata {
    \foreach \val [count = \j] in \row {
      \node at (\j*\cellsize - 0.5*\cellsize, -\i*\cellsize + 0.5*\cellsize) {\val};
    }
  }

  \draw[line width=2pt] (0,1pt) -- (0,-9*\cellsize-1pt);
  \draw[line width=2pt] (3*\cellsize,0) -- (3*\cellsize,-9*\cellsize);
  \draw[line width=2pt] (6*\cellsize,0) -- (6*\cellsize,-9*\cellsize);
  \draw[line width=2pt] (9*\cellsize,1pt) -- (9*\cellsize,-9*\cellsize-1pt);

  \draw[line width=0.4pt] (\cellsize,0) -- (\cellsize,-4*\cellsize);
  \draw[line width=0.4pt] (3*\cellsize,0) -- (3*\cellsize,-4*\cellsize);

  \foreach \y in {0,3,6,9} {
    \draw[line width=2pt] (0,-\y*\cellsize) -- (9*\cellsize,-\y*\cellsize);
  }

  \foreach \y in {1,2,3,4,5,6,7,8,9} {
    \draw[line width=0.4pt] (0,-\y*\cellsize) -- (9*\cellsize,-\y*\cellsize);
  }

  \foreach \x in {1,2,3,4,5,6,7,8,9} {
    \draw[line width=0.4pt] (\cellsize*\x,0) -- (\cellsize*\x,-9*\cellsize);
  }
\end{tikzpicture}
    \caption{An example of an unsolved Sudoku puzzle. 
    The empty cells have to filled in by the player according to the puzzle constraints. 
    The filled cells have a fixed value.}
    \label{fig:sudoku_9_9}
\end{figure}

\subsection{Hopfield neural networks to solve Sudokus}
In an article from 2008 Hopfield proposed a method of solving Sudokus using neural networks \cite{hopfield_searching_2008}, where the encoding exploits a $9 \times 9 \times 9$ matrix of binary neurons $\sigma \in \{0, 1\}$.
Each neuron represents a single digit from $1$ to $9$ for each cell of the Sudoku.
The constraints are set up in such a way that only one of these neurons should be active ($\sigma = 1$) per cell.
All other neurons should be inactive ($\sigma = 0$).
Whichever neuron is active indicates which digit should be filled in the related Sudoku cell.
The \textit{known} cells of the Sudoku are introduced by fixing the states of the nine neurons for the related cells to either $\sigma = 0$ or $\sigma = 1$ to encode the given digit.
All other neurons are initially in the $0$ state and should iteratively be updated to find the solution.
As noted by the author, this approach has limitations for more \textit{difficult} puzzles, or in other worlds puzzles with less \textit{known} digits\footnote{
    We could not find a consensus or hard bound on what categorizes a puzzle as \textit{easy} or \textit{difficult}. 
    The difficulty is therefore empirically estimated.
}.
For \textit{difficult} puzzles the neural network can settle into local minima which do not encode a proper solution.
To overcome this, a \textit{coprocessor} was introduced with inhibitory neurons that prevents the network from settling into an incorrect stable state.
However, in this paper, the Hopfield Neural Network approach \textit{without} the coprocessor will be used as point of comparison.
We decide to only use one-shot dynamics, i.e. a single iteration with HNN, to isolate the effects of the novel phase-encoding and constraint embedding that oscillators allow us to use.

\section{Methods}\label{sec:methods}

\begin{figure}
    \centering
    \setlength{\arrayrulewidth}{1pt}
    \renewcommand{\arraystretch}{1.5}
    \begin{tikzpicture}
  \def\cellsize{1cm}
  \def\tabledata{
    {1,2,3,4},
    {3,4,1,2},
    {4,3,2,1},
    {2,1,4,3},
  }

  \foreach \row [count = \i] in \tabledata {
    \foreach \val [count = \j] in \row {
      \node at (\j*\cellsize - 0.5*\cellsize, -\i*\cellsize + 0.5*\cellsize) {\val};
    }
  }

  \draw[line width=2pt] (0,1pt) -- (0,-4*\cellsize - 1pt);
  \draw[line width=2pt] (2*\cellsize,0) -- (2*\cellsize,-4*\cellsize);
  \draw[line width=2pt] (4*\cellsize,1pt) -- (4*\cellsize,-4*\cellsize - 1pt);

  \draw[line width=0.4pt] (\cellsize,0) -- (\cellsize,-4*\cellsize);
  \draw[line width=0.4pt] (3*\cellsize,0) -- (3*\cellsize,-4*\cellsize);

  \foreach \y in {0,2,4} {
    \draw[line width=2pt] (0,-\y*\cellsize) -- (4*\cellsize,-\y*\cellsize);
  }

  \foreach \y in {1,3} {
    \draw[line width=0.4pt] (0,-\y*\cellsize) -- (4*\cellsize,-\y*\cellsize);
  }
\end{tikzpicture}
    \caption{An example of a solved $4 \times 4$ Sudoku grid. The graph representations of Figs. \ref{fig:subgrid}, \ref{fig:column}, \ref{fig:row} are based on the $4 \times 4$ example.}
    \label{fig:sudoku_4_4}
\end{figure}

We will illustrate how the system maps and solves Sudoku through natural dynamics, using a $4\times4$ Sudoku variant, an example of which is depicted in Fig. \ref{fig:sudoku_4_4}.
This simplified case allows for clearer explanation of the mapping and constraints that are required. 
These scale naturally to the $9\times9$ case, or any general Sudoku size.
For the comparison with HNN, Sudokus of size $4\times4, 9\times9$, and $16\times16$ will be used.

\subsection{Phase mapping}\label{ssec:phase_mapping}
Before doing anything else, it is important to clearly define a correspondence between the phase values of the Kuramoto model and the digits of Sudoku.
This phase mapping will fix a number of points on the unit circle, that corresponds unequivocally to the number of digits of the Sudoku.
The mapping is defined as follows:
\begin{equation}\label{eq:phase_encoding}
{\theta_{\text{digit}}} = 
\frac{2\pi(d - 1)}{n_{\text{d}}} \,\,\, ,
\end{equation} 
where $d$ is the digit to be mapped and $n_{\text{d}}$ is the number of digits, which depends on the size of the Sudoku puzzle.
For example, in the $4\times4$ case the digits are mapped as follows:
\begin{subequations}\label{eq:phase_mapping}
    \begin{align}
    1 &\mapsto 0 \\
    2 &\mapsto \frac{\pi}{2}\\
    3 &\mapsto \pi \\
    4 &\mapsto \frac{3\pi}{2},
    \end{align}
\end{subequations}
where on the left the digit is mapped to the phase value (in radians) on the right.
\begin{figure}
    \centering
    \begin{tikzpicture}
    \draw[thick] (0,0) circle(1.5);

    \draw[->] (-2,0) -- (2,0) node[right] {Re};
    \draw[->] (0,-2) -- (0,2) node[above] {Im};

    \filldraw[black] (1.5, 0) circle(2pt) node[below right] {$1, 0$};
    \filldraw[black] (0, 1.5) circle(2pt) node[above left] {$2, \frac{\pi}{2}$};
    \filldraw[black] (-1.5, 0) circle(2pt) node[below left] {$3, \pi$};
    \filldraw[black] (0, -1.5) circle(2pt) node[below right] {$4, \frac{3\pi}{2}$};

\end{tikzpicture}
    \caption{Digits of a $4\times4$ Sudoku mapped onto the unit circle with their respective phase values.}
    \label{fig:unit_circle}
\end{figure}
Fig. \ref{fig:unit_circle} shows the phase mapping on the unit circle.
After the computation, the final phases are used with the inverse of equation \eqref{eq:phase_encoding} to assign a digit.
This value is rounded to the nearest integer to obtain the final digit for a given cell\footnote{We remark that the phases are mapped using $\mod({\theta_{\text{digit}}, 2 \pi})$ arithmetic.}.

\subsection{Graph topology and weight mapping}\label{ssec:graph_topology_weight_mapping}
In the graph representation for ONNs, one identifies the oscillators as the nodes of a graph and the connections as the edges in a graph.
The graph topology needs be designed in such a way that the natural dynamics of the oscillators solve the Sudoku, depending on which cells or nodes are known and which ones are unknown.
There are three specific cases: First, an edge between two nodes with \textit{known} values, second an edge between two nodes with \textit{unknown} values, and finally an edge between a node with a \textit{known} value and a node with an \textit{unknown} value.
Each of these cases will be explained separately.
Following this, a subgraph for each of the constraints of the Sudoku will be introduced, which are the row, column, and subgrid.

We introduce the following notation for the different matrix interactions, to clearly separate the different types of connection encoded in $J_{ij}^{(a, b)}$, where the indices $i, j$ represent the rows and columns, respectively, and the apices $a, b = \{\textit{known}, \textit{unknown}\}$ indicate which type the row or column is of.
First, for connections between two nodes with \textit{known} values, there will be a bidirectional connection, where the weight $J^{(a, b)}_{ij}$ of the connection is defined by the following relationship:
\begin{equation}
J^{(\text{known}, \text{known})}_{ij} =
\begin{cases}
e^{\ii (\theta_{i} - \theta_{j})}, & \text{if } i \neq j \,\,\, ,\\
0, & \text{if } i = j \,\,\, .
\end{cases}
\end{equation}
Note that the coupling is \textit{Hermitian} in general, which is a consequence of how the phases relate to each other. Second, for connections between two nodes with \textit{unknown} values, there will be a bidirectional connection, where the weight $J_{ij}$ of the connection is set to $-1$.
We choose $-1$ for the interaction strength to maximize the phase difference and repulsion between oscillators.
This approach is employed in analogy with ferromagnetic systems, where negative coupling implies repulsion between interacting elements of the system.
Since \textit{unknown} digits need to be unique among the interacting elements, the value of $J$ that ensures maximal repulsion is therefore $-1$.
The equation for the described connections can be formally written as follows
\begin{equation}
    J^{(\text{unknown, unknown})}_{ij} =
    \begin{cases}
    -1, & \text{if } i \neq j \,\,\, ,\\
    0, & \text{if } i = j \,\,\, .
    \end{cases}
\end{equation}
Finally, for connections between a node with a \textit{known} value and a node with an \textit{unknown} value, there will be a unidirectional connection, with a value of $-1$ towards the \textit{unknown} node.
In this way, an \textit{unknown} oscillator does not affect the \textit{known} oscillator, but the \textit{known} oscillator affects the \textit{unknown} one.
The interactions directed from a known node towards an unknown node, can be written in matrix form as:
\begin{equation}
    J^{(\text{unknown, known})}_{ij} =
    \begin{cases}
    -1, & \text{if } i \neq j \,\,\, ,\\
    0, & \text{if } i = j \,\,\, .
    \end{cases}
\end{equation}
Note that conversely to the previous interactions, this portion of the total matrix is not \textit{Hermitian}, due to the directionality, thus the transpose interaction is set to $0$ as $J^{(\text{known, unknown})}_{ij} = \mathbf{0}$, this is due to the lack  reciprocity.
Given the previously introduced matrices, the one encoding all the interactions will be over all permutations of the labels $(a, b)$
\begin{equation}
    J_{ij} = \sum_{(a, b)} J^{(a, b)}_{ij} \,\,\, .
\end{equation}
\begin{figure}
    \centering
    \begin{tikzpicture}[>=Stealth, scale=1, transform shape]

\tikzset{mynode/.style={circle, draw, minimum size=0.9cm, inner sep=1pt}} 

\tikzset{mynodesplit/.style={circle split, draw, minimum size=0.9cm, inner sep = 3pt}} 

\node[mynode] (n1) at (0,2) {$1$};
\node[mynode] (n3) at (0,0) {$3$};
\node[mynodesplit, fill=lightgray] (n2) at (2.5,2) {
$2$ \nodepart{lower} $4$
};
\node[mynodesplit, fill=lightgray] (n4) at (2.5,0) {
$4$\nodepart{lower} $2$
};

\draw[->,thick] (n1) edge["$-1$"] (n2);
\draw[->,thick] (n1) edge["$-1$", near start, inner sep=0.25mm] (n4);

\draw[->,thick] (n3) edge["$-1$", near start, swap, inner sep=0.25mm] (n2);

\draw[->,thick] (n3) edge["$-1$", swap] (n4);
\draw[<->,thick] (n1) edge["$e^{\pm\ii\pi}$", swap] (n3);
\draw[<->,thick] (n2) edge["$-1$"] (n4);

\end{tikzpicture}
    \caption{Example graph representation of the first Sudoku subgrid, dubbed the subgrid and notated symbolically as $\mathcal{S}$.}
    \label{fig:subgrid}
\end{figure}
\begin{figure}
    \centering
    \begin{tikzpicture}[>=Stealth, scale=1, transform shape]

\tikzset{mynode/.style={circle, draw, minimum size=0.9cm, inner sep=1pt}} 

\tikzset{mynodesplit/.style={circle split, draw, minimum size=0.9cm, inner sep = 3pt}} 

\node[mynode] (n1) at (0,4.5) {$1$};
\node[mynode] (n3) at (0,3.0) {$3$};
\node[mynodesplit, fill=lightgray]   (n4) at (0,1.5)
{
$2$ \nodepart{lower} $4$
};
\node[mynodesplit, fill=lightgray]   (n2) at (0,0.0)
{
$4$ \nodepart{lower} $2$
};

\draw[<->,thick] (n1) edge["$e^{\pm\ii\pi}$"] (n3);
\draw[->,thick] (n3) edge["$-1$"] (n4);
\draw[<->,thick] (n4) edge["$-1$"] (n2);

\draw[->,thick, bend left=60] (n1) edge["$-1$"] (n4);
\draw[->,thick, bend right=40, inner sep=0.3mm] (n3) edge["$-1$", swap] (n2);
\draw[->,thick, bend right=60] (n1) edge["$-1$", swap] (n2);

\end{tikzpicture}
    \caption{Example graph representation of the first Sudoku column, dubbed the column and notated symbolically as $\mathcal{C}$.}
    \label{fig:column}
\end{figure}
\begin{figure}
    \centering
    \begin{tikzpicture}[>=Stealth, scale=1, transform shape]

\tikzset{mynode/.style={circle, draw, minimum size=0.9cm, inner sep=1pt}} 

\tikzset{mynodesplit/.style={circle split, draw, minimum size=0.9cm, inner sep = 3pt}} 

\node[mynode] (n1) at (0,0) {1};
\node[mynode] (n2) at (2,0) {2};
\node[mynodesplit, fill=lightgray] (n3) at (4,0) {
$3$ \nodepart{lower} $4$
};
\node[mynodesplit, fill=lightgray] (n4) at (6,0) {
$4$\nodepart{lower} $3$
};

\draw[<->,thick] (n1) edge["$e^{\pm\ii\frac{\pi}{2}}$"] (n2);
\draw[->,thick] (n2) edge["$-1$"] (n3);
\draw[<->,thick] (n3) edge["$-1$"] (n4);

\draw[->,thick, bend right=30] (n1) edge["$-1$", swap] (n3);
\draw[->,thick, bend left=30] (n2) edge["$-1$"] (n4);
\draw[->,thick, bend left=50] (n1) edge["$-1$"] (n4);

\end{tikzpicture}
    \caption{Example graph representation of the first Sudoku row, dubbed the row and notated symbolically as $\mathcal{R}$.}
    \label{fig:row}
\end{figure}
For clarity, example graphs for a subgrid, a row, and a column, are given in Figures~\ref{fig:subgrid},~\ref{fig:column} and~\ref{fig:row}.
One can symbolically subdivide the general Kuramoto dynamics given in Eq. \eqref{eq:kuramoto-model} into one Ordinary Differential Equation (ODE) for each puzzle constraint.
To make the notation for constraints less cumbersome, we introduce a set notation, where the nodes involved in one constraint are included in the specific set.
For each oscillator $i$ these sets are the subgrid $\mathcal{S}_{i}$, row $\mathcal{R}_{i}$, and column $\mathcal{C}_{i}$.
The union of these sets is the set $\Omega_i$ of all constraints for that oscillator.
Consequently, the subdivided Kuramoto dynamics are given by the following equations:
\begin{subequations}\label{eq:constraint-kuramoto-equations}
    \begin{equation}\label{eq:subgrid}
    \dot{\theta}_{i}^{(\mathcal{S})}= - \sum_{j \in \mathcal{S}_{i}} J^{(\mathcal{S}_{i})}_{ij} \sin(\theta_{i} - \theta_{j})\,,
    \end{equation}
    
    \begin{equation}\label{eq:row}
    \dot{\theta}_{i}^{(\mathcal{R})} = - \sum_{j \in \mathcal{R}_{i}} J^{(\mathcal{R}_{i})}_{ij} \sin(\theta_{i} - \theta_{j})\,,
    \end{equation}
    
    \begin{equation}\label{eq:column}
    \dot{\theta}_{i}^{(\mathcal{C})}= - \sum_{j \in \mathcal{C}_{i}} J^{(\mathcal{C}_{i})}_{ij} \sin(\theta_{i} - \theta_{j})\,,
    \end{equation}
\end{subequations}
and the equation that defines the union is given by
\begin{equation}\label{eq:omegai}
\Omega_{i} = \mathcal{S}_{i} \cup \mathcal{R}_{i} \cup \mathcal{C}_{i}\,.
\end{equation}
Finally, the union of all oscillator sets defines the complete puzzle as:
\begin{equation}
\Omega = \bigcup_{i = 1}^{N} \Omega_{i} \, ,
\label{eq:omega}
\end{equation}
where $N$ is the number of elements in the Sudoku, i.e. in a $4\times4$  Sudoku $N=16$, and in a $9 \times 9$ Sudoku $N = 81$, in general $ N = n^{2}_{\text{d}}$\footnote{
Note that we introduced $N$ before as the number of oscillators in equation \ref{eq:kuramoto-model}. Due to the mapping these numbers match exactly, so we have decided to reuse the symbol.
}.

As a consequence, the equation for the total natural dynamics is then the linear combination of equations \eqref{eq:subgrid}, \eqref{eq:row}, and \eqref{eq:column}, which can be succinctly written as:
\begin{equation}
\dot{\theta}_{i} = 
- \sum_{j \in \Omega_{i}} J_{ij} \sin(\theta_{i} - \theta_{j}) \,\,\, , \,\,\, \forall i \in \{1, \dots, N \}.
\label{eq:total}
\end{equation}
The capability of ONNs to encode information is fully exploited, where instead of using multiple neurons per cell, as it has been done in \cite{hopfield_searching_2008, mladenov2011solving}, one cell is directly represented by one oscillator.
By using only one oscillator per cell, the network already becomes less complex than the network in \cite{hopfield_searching_2008, mladenov2011solving}.
In general, this would mean that for an $n_{\text{d}}$-valued Sudoku, the spatial complexity in the two cases would be $\mathcal{O}(n^{2}_{\text{d}})$ for the ONN, and $\mathcal{O}(n^{3}_{\text{d}})$ for the HNN.
To represent the network, one must define a weight matrix to generate the connections among nodes \footnote{
To remain agnostic between the two paradigms, we use the term node. Normally one would say neuron for HNNs and oscillator (phase) for ONNs.
}.
The weight matrix sizes are $n^{4}_{\text{d}}$ for the ONN, and $n^{6}_{\text{d}}$ for the HNN version \cite{hopfield_searching_2008, mladenov2011solving}.
Specifically, for a standard $9\times9$ Sudoku, this work requires $81$ nodes and a $81\times81$ weight matrix, whereas \cite{hopfield_searching_2008, mladenov2011solving} needs $729$ nodes and a $729 \times 729$ weight matrix.
A lower amount of nodes, and thus a lower number of connections, allows for an easier implementation in hardware and lessens the computational load to be simulated.

\subsection{Order parameter}\label{ssec:order_parameter}
Although it is customary in the context of ONNs to link the phase dynamics to a gradient descent of a Hamiltonian \cite{todri-sanial_computing_2024}, this formulation is directly applicable only if the system possesses a symmetric coupling matrix $J_{ij}$.
Since this is not the case in our mapping for Sudoku, to quantify the closeness of the system to a solution, we exploit the \textit{synchronization order parameter} $\rho_{ n_{\text{d}} }$ of order $n_{\text{d}}$ defined as
\begin{equation}
    \rho_{n_{\text{d}}} := \bigg| \dfrac{1}{N} \sum_{i = 1}^{N} e^{\ii n_{\text{d}} \theta_{i} } \bigg| \,\,\, , \,\,\, \rho_{n_{\text{d}}} \in [0, 1] \,\,\, ,
\end{equation}
which quantifies the degree of synchronization of the network as introduced by Daido \cite{daido_order_function_1992, DAIDO1993394, daido_generic_scaling_1994, DAIDO199624}.
The choice of the $n_{\text{d}}$-th order parameter, stems from the fact that it quantifies the synchronization into $n_{\text{d}}$ populations by construction.
A value of $\rho_{n_{\text{d}}} \simeq 0$ implies a mostly unsynchronized network, whereas a value of $\rho_{n_{\text{d}}} \simeq 1$ implies a synchronized network.
Due to the phase mapping introduced in section \ref{ssec:phase_mapping}, the amount of populations to synchronize to coincides with the number of distinct digits of the puzzle.
The \textit{order parameter} $\rho_{n_{\text{d}}}$ is a measure of how close the phases have evolved to a value that maps to the digit of the puzzle.
The closer $\rho_{n_{\text{d}}}$ is to $1$, the more likely the system has discretized to one of the digits encoded on the unit circle.
This is a \textit{necessary condition} to have a correct solution, however, this is not a \textit{sufficient condition}.
In other words, this implies that the system can fully discretize while encoding a wrong solution.
Conversely, it is possible for the system to not be completely discretized to the encoded digits, while still providing a correct solution of the Sudoku, due to the rounding procedure discussed in section \ref{ssec:phase_mapping}.

\subsection{Time-to-settle}\label{ssec:time_to_settle}
The \textit{Time-to-Settle} (TTS) is an important metric for assessing the performance of an ONN in terms of runtime. Typically, this time is normalized to oscillation cycles, a frequency-independent metric. This is especially useful since the utilized ONN simulation operates using arbitrary time units. To obtain the TTS, a stability metric is needed.
We will reuse the order parameter $\rho$, introduced in section \ref{ssec:order_parameter} for this.
First, the time trace of the order parameter of the phases is taken.
From this, the change of the order parameter is computed between each time step.
Then a moving average with a window size of one half oscillation cycle, is taken of the change of the order parameter. 
If the average change in the window is less than a threshold of $10^{-5}$, the ONN is considered settled.
Then the lowest time index where this is true is taken as the TTS for that run.
The TTS is stored for every run, then for each combination of Sudoku size and unknown ratio the average and standard deviation are computed.
With enough samples, this should provide a good estimate of the time an ONN needs to solve a given Sudoku puzzle.
We do not report TTS for Sudoku sizes or unknown ratios with empirically low solution quality ($<0.001$ ratio of correctly solved solutions).

\subsection{Simulations}\label{ssec:simulations}
\begin{figure}
    \centering
    \includegraphics[width=\linewidth]{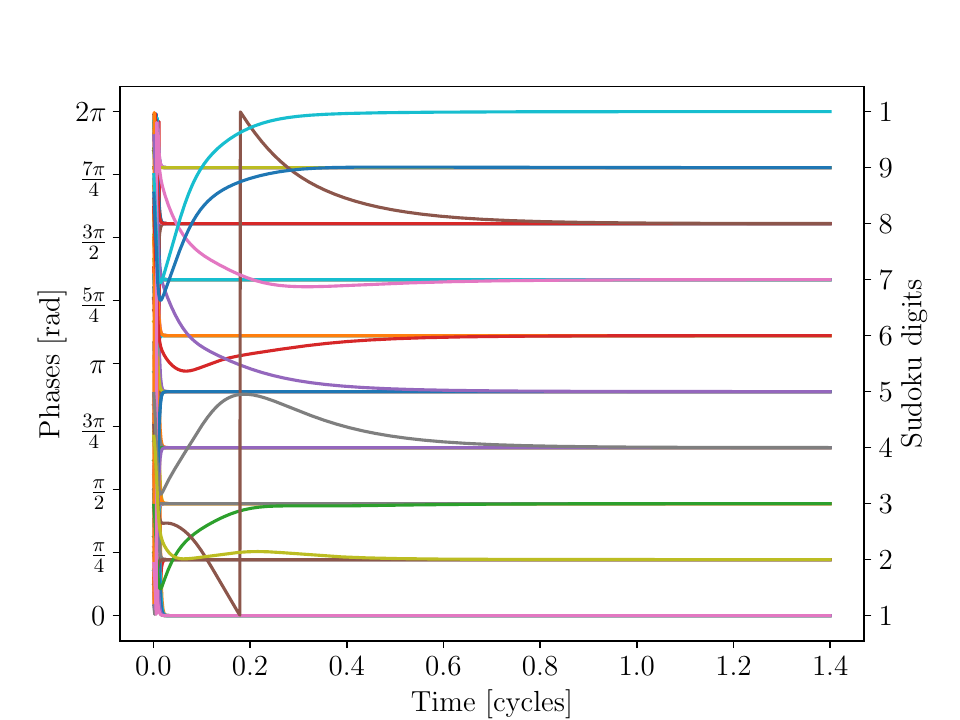}
    \caption{Phase evolution of an ONN correctly solving a $9\times9$ Sudoku with 10 unknown cells.
    The cycles do not possess any physical time unit, and should therefore be regarded to as arbitrary units.
    The Sudoku digits have $\mod(\cdot, 2 \pi)$ periodicity, thus the phases that converge to $2 \pi$ are mapped back to $1$.}
    \label{fig:phase-evolution}
\end{figure}

We simulate the Kuramoto model in Python, using the Runge-Kutta solver of order 4.
The first\footnote{
    First when reading the unsolved puzzle from left to right, top to bottom.
} \textit{known} oscillator is taken as the reference and all phases are computed relative to that\footnote{
    Note that any oscillator can be the reference oscillator, even an unknown oscillator. This will simply rotate the reference frame and therefore the digits, but the relative difference between each oscillators remains the same.
}.
For illustration, Fig. \ref{fig:phase-evolution} shows a complete time trace of a randomized $9\times9$ Sudoku puzzle.
In addition to solving Sudoku with ONNs, we compare the results to a Hopfield network architecture, based on the articles \cite{hopfield_searching_2008, mladenov2011solving}, to provide a reference point and highlight the differences of the proposed approach.
The HNN is classically used iteratively, but since ONNs employ a "one-shot" kind of approach, namely once the initial conditions are set the system freely runs, we choose to use only one iteration for the HNN.
A total of ten fully solved puzzles per Sudoku size were generated using Py-Sudoku \cite{Sieu2019_PySudoku}.
Simulations were conducted for each of these puzzles with the ratio of unknowns to total cells from $0.1$ to $0.5$ in increments of $0.1$.
For brevity we introduce the symbol $\eta$ as follows:
\begin{equation}
    \eta = \frac{N_\text{unknown}}{N},
\end{equation}
where $N_\text{unknown}$ is the number of cells with an \textit{unknown} value.
For each specified ratio of unknown cells, that number of values, rounded up, was randomly removed from the fully solved Sudoku puzzle to generate the puzzle input.
This removal process was performed 100 times per puzzle using a randomized masking algorithm to create randomized Sudoku puzzles while ensuring that a valid solution remained.
For each case, 10 independent simulation runs were executed, each with randomly initialized phases sampled from a uniform distribution in the range $[0, 2\pi)$.
In short, for each Sudoku size and difficulty, $1000$ different puzzles are used, each of which are sampled $10$ times with random initial phases, or in total 10,000 samples per data point.
This amount should provide ample data to draw conclusions.

\section{Results}\label{sec:results}

\begin{figure}
    \centering
    \includegraphics[width=\linewidth]{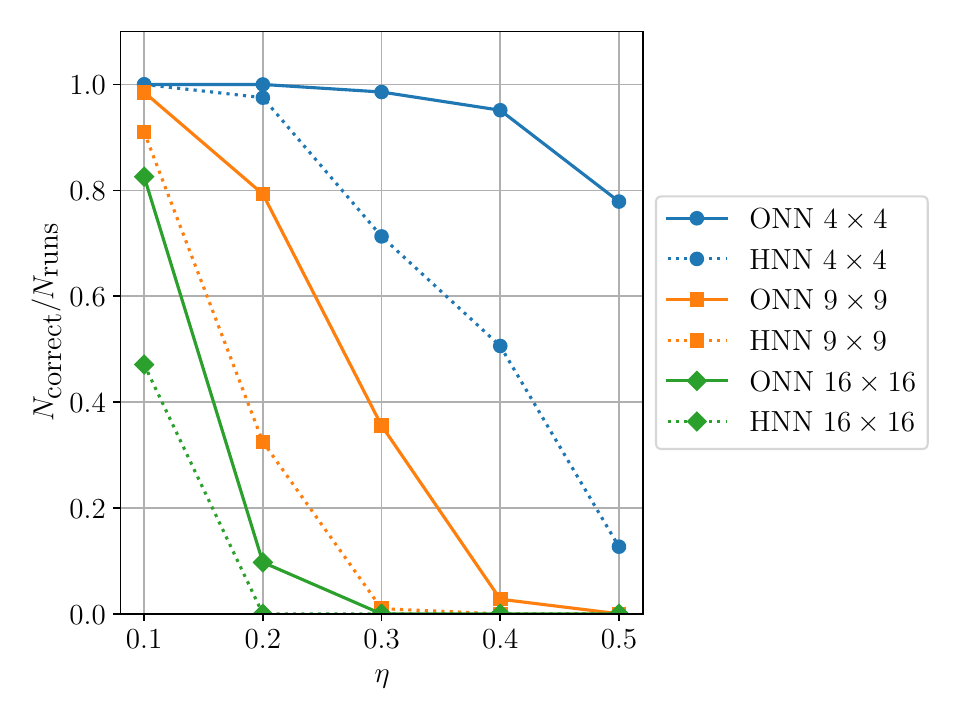}
    \caption{Comparison of ONN and HNN performance across different Sudoku sizes and $\eta$.}
    \label{fig:onn_vs_hopfield}
\end{figure}

Fig.~\ref{fig:onn_vs_hopfield} and Table \ref{tab:onn_vs_hopfield} show the results of the benchmarks.
Two general trends are observed:
First, as $\eta$ increases, the ratio of correct results decreases.
Second, as the Sudoku size increases, the ratio of correct results decreases.
When looking at only the results for the $4\times4$ Sudoku, it is seen that both the ONN and HNN achieve 100\% correct results at $\eta=0.1$.
For the ONN this gradually drops off to just below 80\% at $\eta=0.5$.
The performance of the HNN has a faster decline from $\eta=0.2$ and drops to below 20\% for the \textit{most difficult} puzzles.
For the standard Sudoku size of $9\times9$, the ONN can solve 98\% of the \textit{easiest} puzzles, which declines until almost no puzzles can be solved anymore at $\eta=0.4$.
The HNN approach starts out lower at 91\% for the \textit{easiest} puzzles and declines faster where at $\eta=0.3$ essentially no puzzles are solved.
At $16\times16$ the ONN approach solves 82\% and 10\% at $\eta=0.1$, and $\eta=0.2$ respectively.
No puzzles are solved above these ratios.
For this size, the HNN solves 47\% at $\eta=0.1$, and no puzzles at all the other ratios.
To summarize, across the whole dataset, the ONN consistently outperformed the Hopfield network.
The ONN not only solves more puzzles on average, but shows a slower decline in performance as the puzzle difficulty increases.

\begin{table}
    \centering
    \caption{Ratio of correctly solved Sudoku puzzles at different scales and $\eta$.}
    \label{tab:onn_vs_hopfield}
    \begin{tabular}{|c|c|c|c|c|c|}
        \hline
        $\eta$ & 0.1 & 0.2 & 0.3 & 0.4 & 0.5 \\ \hline \hline
        \multicolumn{6}{|c|}{$4\times4$} \\ \hline
        ONN & 1.000 & 1.000 & 0.986 & 0.951 & 0.779 \\
        HNN & 1.000 & 0.975 & 0.713 & 0.506 & 0.127 \\ \hline \hline
        \multicolumn{6}{|c|}{$9\times9$} \\ \hline
        ONN & 0.985 & 0.793 & 0.356 & 0.028 & 0.000 \\
        HNN & 0.910 & 0.325 & 0.010 & 0.000 & 0.000 \\ \hline \hline
        \multicolumn{6}{|c|}{$16\times16$} \\ \hline
        ONN & 0.826 & 0.097 & 0.000 & 0.000 & 0.000 \\
        HNN & 0.471 & 0.000 & 0.000 & 0.000 & 0.000 \\ \hline
    \end{tabular}
\end{table}

In Table \ref{tab:runtimes} the average time to settle in oscillation cycles can be seen for each Sudoku size and difficulty.
Note that the cycles are normalized to a period of $2\pi$ and measured in arbitrary units.
Larger networks require more time to settle and puzzles with more unknown cells require more time to settle.
In other words, there is a correlation between total puzzle complexity and computation time.
Nevertheless, the time-to-settle is measured at most in tens of cycles.
To illustrate, real-world oscillators with a frequency around a gigahertz would compute a solution in tens of nanoseconds.
\renewcommand{\arraystretch}{1.5} 
\begin{table}
    \centering
    \caption{Average time to settle in oscillation cycles for ONN at different scales and $\eta$.} 
    \begin{tabular}{|c|c|c|c|c|c|}
        \hline
        $\eta$ & 0.1 & 0.2 & 0.3 & 0.4 & 0.5 \\ \hline \hline
        {$t_{4 \times 4}[\text{cycles}]$\footnotemark[1]} & \makecell{0.36\\$\pm0.16$} & \makecell{0.48\\$\pm0.25$} & \makecell{0.82\\$\pm0.38$} & \makecell{1.00\\$\pm0.43$} & \makecell{1.39\\$\pm0.61$} \\
        {$t_{9 \times 9}[\text{cycles}]$\footnotemark[1]} & \makecell{0.70\\$\pm0.41$} & \makecell{1.85\\$\pm1.26$} & \makecell{4.04\\$\pm2.48$} & n/a & n/a \\
        {$t_{16 \times 16}[\text{cycles}]$\footnotemark[1]} & \makecell{1.74\\$\pm1.04$} & \makecell{5.47\\$\pm2.51$} & n/a & n/a & n/a \\ \hline
    \end{tabular}
    \label{tab:runtimes}
    \vspace{5pt}
    \footnotesize{$1$: The cycles are measured in arbitrary units, normalized to a period of $2 \pi$. \\
    The values $\pm$ are the standard deviations.}
\end{table}

\begin{figure}
    \centering
    \includegraphics[width=\linewidth]{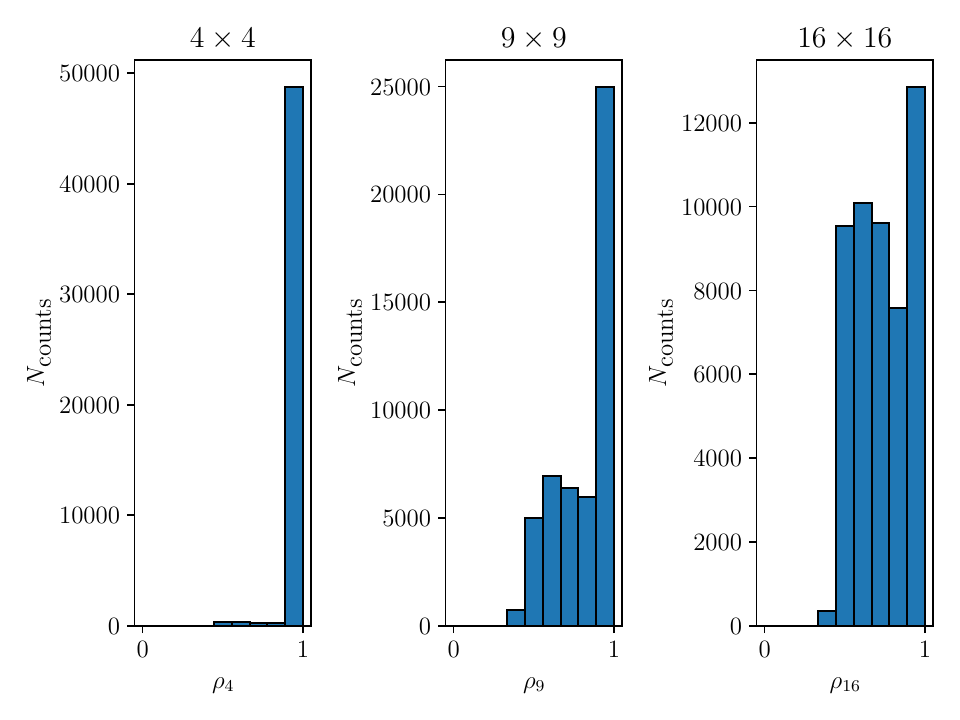}
    \caption{
        The three histograms represent the counts at different values of the order parameter $\rho_{n_{\text{d}}}$ introduced in subsection \ref{ssec:order_parameter}. 
        These are subdivided by Sudoku size, and all the unknown-to-known ratios are aggregated, regardless of the known-to-unknown ratio. 
        In the $4 \times 4$ grid most values are concentrated around $1$, namely the system mostly solves the puzzles. 
        In the $9 \times 9$, the synchronization still dominates, whereas in the $16 \times 16$ the system struggles to reach a state that is synchronized.
    }\label{fig:unnormalized_counts_order_parameter}
\end{figure}

\renewcommand{\arraystretch}{1}
As introduced in subsection \ref{ssec:order_parameter}, the order parameter $\rho_{n_{\text{d}}}$ is an indicator of how far the system is from being fully synchronized to the states encoding the digits.
We use a histogram with values for $\rho_{n_{\text{d}}}$ on $[0, 1]$, which is divided up into ten bins.
The results are shown in Fig. \ref{fig:unnormalized_counts_order_parameter} for the different Sudoku grid sizes.
Bins that are farther farther from the bin containing $\rho_{n_{\text{d}}} = 1$ correspond to lower synchronization of the network.
In the aggregated histograms from Fig. \ref{fig:unnormalized_counts_order_parameter} the bins with lower values of $\rho_{n_{\text{d}}}$ are correlated with wrong solutions.
%
In Fig. \ref{fig:unnormalized_counts_order_parameter} the histograms reach lower values with increasing size of the Sudoku grid more often.
This means, that the larger the Sudoku, the more difficult for the system it is to settle to correctly discretized values.
One can observe, together with the results for the ONN reported in Fig. \ref{fig:onn_vs_hopfield}, how the $4 \times 4$ has the highest amount of correctly solved puzzles, as well as the highest amount of synchronized oscillators as shown in Fig. \ref{fig:unnormalized_counts_order_parameter}.
Increasing the number of unknowns as well as network size Fig. \ref{fig:onn_vs_hopfield} displays a decrease in performance, which in conjunction with Fig. \ref{fig:unnormalized_counts_order_parameter}, it is possible to observe an empirical correlation among network size, number of unknowns, and synchronization of the network of coupled oscillators.
%

\section{Discussion}\label{sec:discussion}
The results presented in section \ref{sec:results} show that ONNs are capable of solving Sudoku puzzles with performance varying depending on the puzzle size and difficulty.
For $4 \times 4$ puzzles, slightly less than 80\% of puzzles were still solved with half of the cells empty initially.
At $9 \times 9$ and $16 \times 16$ similar solution ratios are seen at $\eta=0.2$ and $\eta=0.1$, respectively.
This indicates that not only the Sudoku difficulty matters, but also the Sudoku size for the quality of the results.
We observe that the ONN continues to provide some correct solutions even in cases where a baseline Hopfield network fails entirely.
At larger sizes, it is possible that unexpected stable points form that do not encode proper solutions.
By adding an extra penalty term, these stable points could potentially be removed.
The way to modify the dynamics, would be to include a term that explicitly enforces a penalty on the dynamical behavior of the system as long as not all constraints are satisfied and the steady-states are not within an accepted synchronization tolerance, will be considered for future work.

Table \ref{tab:comparison_onn_hnn} highlights the qualitative properties of the two paradigms.
The key point is the much lower order of scaling for both the number of neurons and number of connections for the ONN approach when compared to the HNN approach.
This is made possible by the multi-digit phase encoding that an ONN supports.
The TTS is difficult to compare one-to-one between the two approaches, since an HNN works using discrete update steps and ONNs operate with continuous dynamics.
In the HNN paradigm, there is no physical meaning to the time steps, whereas solving an equation through ONN dynamics with the Kuramoto ODE \eqref{eq:kuramoto-model}, time has a physical (albeit in arbitrary units) meaning.
\begin{table}
    \centering
    \caption{Comparison of ONN and HNN properties for solving Sudoku}
    \label{tab:comparison_onn_hnn}
    \begin{tabular}{|c|c|c|}
        \hline
        & \textbf{ONN} & \textbf{HNN}
        \\ \hline\hline
        \textit{Encoding} & Relative phases & Binary/Spins 
        \\ \hline
        \textit{Mapping/Architecture} & Natural dynamics & \textit{Neural Network Coprocessor} 
        \\ \hline
        \textit{Steps} & One-shot & Iterative 
        \\ \hline
        \textit{Time-to-Settle} & Problem dependent ($\lessapprox 10$ cycles)  & 1 iteration
        \\ \hline
        \textit{\# Neurons} & $\mathcal{O}(n_\text{d}^2)$   & $\mathcal{O}(n_\text{d}^3)$ 
        \\ \hline
        \textit{\# Connections} & $\mathcal{O}(n_{\text{d}}^4)$   & $\mathcal{O}(n_\text{d}^6)$ 
        \\ \hline
    \end{tabular}
\end{table}
\section{Conclusions}\label{sec:conclusions}
This paper explored the use of Oscillatory Neural Networks (ONNs) for solving Sudoku puzzles as a proof of concept, by leveraging their phase-based dynamics to enforce constraints according to the given rules.
The novelty of the approach relies on engineering on embedding the sudoku problem and its constraints, rather than relying on training methods or iterative approaches.
As a result, this choice is exemplary of exploiting the physical dynamics of the system to solve the problem, rather than employing pure algorithmics.
%
The ONN architecture exploits the constraints by enforcing them directly into the total coupling matrix $J_{ij}$ between oscillators, and having fixed known values to guide synchronization to the right solution.  
Simulation results show how the ONN paradigm performs well for $4\times4$ puzzles with up to half unknown digits, correctly solving around $78\%$ of the puzzles. 
Although performance decreased with more \textit{difficult} and larger puzzles, the ONN continued to produce some correct results even for a $16 \times 16$ Sudoku with 20\% of unknown cells. 
%
We have also shown that the quality of the network synchronization is related to the size of the puzzle.
Compared to a baseline binary HNN architecture, the ONN showed greater robustness and adaptability, especially in moderately difficult puzzles, showing its potential for constraint satisfaction problems.
Overall, the study shows that ONNs offer a promising alternative for solving constraint satisfaction problems such as Sudoku and its direct link with the natural dynamics of the underlying system dynamics, in stark contrast to traditional algorithmics.

\section*{Declarations}
This work has received funding from the European Research Council ERC THERMODON project under grant agreement No. 101125031 and Dutch Research Council's AiNed Fellowship research program, AI-on-ONN project under grant agreement No. NGF.1607.22.016.

This work encompasses the graduation project of Stefan Porfir, where principal supervision was performed by Bram F. Haverkort and Federico Sbravati.
The main conceptualization was done by Bram F. Haverkort, Federico Sbravati, and Professor Aida Todri-Sanial.
Professor Aida Todri-Sanial co-supervised and is the principal investigator of Bram F. Haverkort and Federico Sbravati.
Professor Aida Todri-Sanial is responsible for all funding relations.

\bibliography{references}  

\end{document}